\documentclass[a4paper,11pt]{article}
\usepackage{pos}

\usepackage{graphicx}
\usepackage[makeroom]{cancel}  
\usepackage{wrapfig}
\usepackage{subcaption}

\usepackage{xcolor}
\colorlet{alert}{red!60!black}
\colorlet{example}{green!60!black}
\colorlet{structure}{blue!60!black}

\usepackage{hyperref}
\hypersetup{
    colorlinks,
    linkcolor={red!50!black},
    citecolor={blue!50!black},
    urlcolor={blue!80!black}
}

\newcommand{\mnote}[1]{}                   

\newcommand{\ind}[1]{\rm\scriptscriptstyle #1}

\def\lsim{\mathrel{\rlap{\lower4pt\hbox{\hskip1pt$\sim$}}
    \raise1pt\hbox{$<$}}}                
\def\gsim{\mathrel{\rlap{\lower4pt\hbox{\hskip1pt$\sim$}}
    \raise1pt\hbox{$>$}}}                

\begin{document}


\title{
\vspace*{-0.75cm}
\begin{minipage}{\textwidth}
\begin{flushright}
\texttt{\footnotesize
PoS(LATTICE2024)323  \\
ADP-25-2/T1264       \\
DESY-25-011          \\
Liverpool LTH 1391   \\
}
\end{flushright}
\end{minipage}\\[15pt]
\vspace*{+0.75cm}
        Renormalisation Group Equations for 2+1 clover fermions}

\ShortTitle{RG equations for 2+1 clover fermions}

\author[a]{K.~U.~Can}
\author*[b,1]{R.~Horsley}
\author[c]{P.~E.~L.~Rakow}
\author[d]{G.~Schierholz}
\author[e]{H.~St\"uben}
\author[a]{R.~D.~Young}
\author[a]{J.~M.~Zanotti}

\affiliation[a]{CSSM, Department of Physics,
               University of Adelaide, Adelaide SA 5005, Australia}
\affiliation[b]{School of Physics and Astronomy, University of Edinburgh,
                Edinburgh EH9 3FD, UK}
\affiliation[c]{Theoretical Physics Division,
                Department of Mathematical Sciences,
                University of Liverpool, \\
                Liverpool L69 3BX, UK}
\affiliation[d]{Deutsches Elektronen-Synchrotron DESY,
                Notkestr. 85, 22607 Hamburg, Germany}
\affiliation[e]{Universit\"at Hamburg, Regionales Rechenzentrum,
                20146 Hamburg, Germany}

\note{For the QCDSF Collaboration}

\emailAdd{rhorsley@ph.ed.ac.uk}

\abstract{Many lattice QCD simulations now have many lattice spacings 
          available, and it is of interest to investigate how they scale.
          In this talk we first derive renormalisation group equations
          appropriate for 2+1 clover fermions. This is then used together
          with pion mass and gradient flow results at five lattice spacings
          to study scaling.}

\FullConference{%
   The 41st International Symposium on Lattice Field Theory (LATTICE2024)\\
   28 July - 3 August 2024\\
   Liverpool, UK\\}

\maketitle


\section{Introduction}
\label{introduction}


One of the major difficulties in lattice gauge theories is the extrapolation
to the continuum limit. Usually individual fits at a particular lattice 
spacing are made to, e.g.\ a hadron mass or matrix element, followed by an 
extrapolation to the continuum limit (usually in $a^2$). This is a noisy 
procedure, involving both the pre-determined $a$ values and also the 
individual values of the mass or matrix element. To improve the situation often
a smoothing procedure is used, for example a joint extrapolation
involving both the lattice spacing and mass or matrix element simultaneously.
Theoretically this is often not so well justified. A better motivated
procedure suggested here is to use the Renormalisation Group (RG) equations 
to help to achieve this. In this contribution we attempt a first step 
by deriving and solving the RG equations for the lattice spacing.


\section{The RG equation}
\label{RG_eqn}


Scaling means that we have lines of constant physics (e.g.\ constant mass 
ratios) passing through our parameter space. On one of these lines we 
can measure the relative rate at which lattice spacing changes by 
monitoring the rate at which correlation lengths change as we move along 
the trajectory. Scaling (up to $O(a^2)$ violations) are regions where 
$a\Lambda^{\ind lat} \ll 1$ and $am_q \ll 1$. 
For Wilson type fermions considered here we have in $g_0$, $\kappa_q$ space
\begin{eqnarray}
   \left.  a {\partial \over \partial a} \right|_{\ind physics}
      = U(g_0, \{\kappa \}) 
               \left. \frac{ \partial}{\partial g_0} \right|_\kappa
        + \sum_q V_q(g_0,\{\kappa \})
               \left. {\partial \over \partial \kappa_q} \right|_{g_0} \,.
\end{eqnarray}
Re-writing this in terms of $m_q^{\ind lat} \equiv am_q$ where
\begin{eqnarray}
   a m_q \equiv {1 \over 2} \left( {1 \over \kappa_q}
                                         - {1 \over \kappa_{0c}(g_0)}
                            \right) \,,
\end{eqnarray}
(with $\kappa_0$ the hopping parameter on the $SU(n_f)$-flavour symmetric
line, with critical point at the chiral limit $\kappa_{0c}$) and then 
Taylor expanding $U$ and $V_q$ gives
\begin{eqnarray}
   a \left. {\partial \over \partial a} \right|_{\ind physics}
      &=& B_0(g_0) \left. {\partial \over \partial g_0} \right|_{\{am\}}
          + B_1(g_0) a \bar{m} 
                   \left. {\partial \over \partial g_0} \right|_{\{am\}}
                                                           \nonumber  \\
      & & + G_0(g_0) \sum_q a m_q  
                   \left. {\partial \over \partial a m_q} \right|_{g_0}
          + H_0(g_0)  a \bar{m} \sum_q
                   \left. {\partial \over \partial a m_q} \right|_{g_0} 
                                                           \nonumber  \\
      & & + G_1(g_0) a \bar{m} \sum_q a m_q 
                   \left. {\partial \over \partial a m_q} \right|_{g_0} 
          + G_2(g_0) \sum_q (a m_q)^2 
                   \left. {\partial \over \partial a m_q} \right|_{g_0} 
                                                           \nonumber  \\
      & & + H_1(g_0) (a \bar{m})^2  \sum_q  
                   \left. {\partial \over \partial a m_q} \right|_{g_0} 
          + H_2(g_0)  a^2 \widebar{m^2} \sum_q 
                   \left. {\partial \over \partial a m_q} \right|_{g_0} \,,
\end{eqnarray}
where  
\begin{eqnarray}
   a \bar{m} \equiv {1 \over n_f} \sum_q a m_q \,,
   \qquad 
   a^2 \widebar{m^2} \equiv {1 \over n_f} \sum_q a^2 m_q^2 \,.
\end{eqnarray}
The $B_0$, $G_0$, $H_0$ coefficients are usually referred to as
`renormalisation', while the $B_1$, $G_1$, $G_2$, $H_1$, $H_2$  coefficients
are the `improvement' terms, so in principle this treats both the 
renormalisation and improvement terms in a unified fashion. For chiral
fermions all the coefficients vanish, except for $B_0$ and $G_0$.

It is sufficient to only consider here the leading terms (ie $O(am_q)$),
although the higher order terms can be considered, \cite{qcdsf25}.
The leading functions are related to the usual renormalisation coefficients
\begin{eqnarray}
   B_0(g_0) = - \beta(g_0) \,, \quad
   G_0(g_0) = 1 - \gamma_m^{\ind NS}(g_0) \,, \quad
   H_0(g_0) = \gamma_m^{\ind NS}(g_0) - \gamma_m^{\ind S}(g_0) \,,
\label{PT_funs}
\end{eqnarray}
with
\begin{eqnarray}
   \begin{aligned}
      \beta(g_0) &= - b_0g_0^3 - b_1g_0^5 
                               - b_2^{\ind lat}g_0^7 + O(g_0^9) \,, \\
      B_1(g_0)   &= \phantom{-} b^{\ind lat}_{10}g_0^{3} + O(g_0^5) \,,
   \end{aligned}
   \qquad
   \begin{aligned}
      \gamma_m^{\ind NS}(g_0) 
              &= d_0 g_0^2 + d_1^{\ind NS\,lat} g_0^4 + O(g_0^6) \,, \\
      \gamma_m^{\ind S}(g_0) 
              &= d_0 g_0^2 + d_1^{\ind S\,lat} g_0^4 + O(g_0^6) \,,
   \end{aligned}
\end{eqnarray}
where $b_0$, $b_1$ and $d_0$ are known universal coefficients.


\subsection{Renormalised quark masses}


For Wilson-like fermions, which have no chiral symmetry, the singlet and
non-singlet quark mass pieces evolve differently
\begin{eqnarray}
   \bar{m}^{\ind rgi} = a\bar{m}\,u(g_0)\,, \qquad
   m_q^{\ind rgi} - \bar{m}^{\ind rgi}  = (am_q - a\bar{m})\,v(g_0) \,.
\label{mbarrgi_mbar}
\end{eqnarray}
For simplicity we consider the Renormalised Group Invariant (RGI)
quark mass, alternatively the same equations hold for the quark mass in a 
given scheme at a given scale. Also to simplify the notation somewhat we have 
absorbed $\Lambda^{\ind lat}$ into the definition of $\bar{m}^{\ind rgi}$, 
i.e.\ we measure the quark mass in units of $\Lambda^{\ind lat}$.
Note, in particular, that $\bar{m}^{\ind rgi} \propto a\bar{m}$.

We now solve the RG equation
\begin{eqnarray}
   a \left. {\partial \over \partial a} \right|_{\ind physics} m_q^{\ind rgi} 
      = 0 \,,
\end{eqnarray}
to give
\begin{eqnarray}
   u(g_0) = C(g_0) 
              \exp\left[ \int_0^{g_0} d\xi 
               \left\{ {1 \over b_0 \xi^3} - {b_1 + b_0 d_0 \over b_0^2 \xi} 
                      - {G_0(\xi)+H_0(\xi) \over B_0(\xi)} \right\} 
               \right] \,,
\end{eqnarray}
and
\begin{eqnarray}
  v(g_0) = C(g_0) 
             \exp\left[ \int_0^{g_0} d\xi 
              \left\{ {1 \over b_0 \xi^3} - {b_1 + b_0 d_0 \over b_0^2 \xi} 
                     - {G_0(\xi) \over B_0(\xi)} \right\} \right] \,,
\end{eqnarray}
where
\begin{eqnarray}
   C(g_0) 
      = \Lambda^{\ind lat} 
           \left[ 2 b_0 g_0^2 \right]^{d_0 \over 2 b_0}
                  \left[ b_0 g_0^2 \right]^{b_1 \over 2 b_0^2}
                  \exp\left[ \frac{1}{2 b_0 g_0^2} \right] \,.
\end{eqnarray}
These are the usual type of equations, arranged so that there is no
singularity in the integral as $g_0 \to 0$. Again, for simplicity,
as for $\bar{m}^{\ind rgi}$ we shall in future measure $u$ (and $v$) in units
of $\Lambda^{\ind lat}$, so $u \to u/\Lambda^{\ind lat}$. Similar equations 
(with appropriate $\gamma$s) also hold for the renormalisation of operators 
$m^{\ind rgi} \to O^{\ind rgi}$. However we shall not consider this further here.


\subsection{Coupling constant}


Rather than considering the renormalised coupling, we consider here the
relation between the lattice spacing and lattice coupling. There are
two possibilities. First we consider changing lattice spacing 
(i.e.\ including $a\bar{m}$ in the definition, but keeping the coupling 
constant). Setting
\begin{eqnarray}
   a = s(g_0) \left\{ 1 + a \bar{m}\, r(g_0) \right\} \,,
\label{a_definition}
\end{eqnarray}
and solving
\begin{eqnarray}
   a \left. {\partial \over \partial a} \right|_{\ind physics} a = a \,,
\end{eqnarray}
gives
\begin{eqnarray}  
   s(g_0)
     &=& A(g_0) \exp\left( \int^{g_0}_0 d\xi \left\{
                  {1 \over B_0(\xi)} - {1 \over b_0 \xi^3}  
                  + {b_1 \over b_0^2 \xi} \right\} \right) \,,
                                                         \label{sg0}  \\
   r(g_0) 
     &=& - u(g_0) \int_0^{g_0} d\xi {B_1(\xi) \over u(\xi) B_0^2(\xi)}
           \,\,\,\underbrace{=}_{g_0^2 \to 0} \,\,\,
               -{b_{10}^{\ind lat} \over b_0} g_0^2 + O(g_0^4) \,,
\end{eqnarray}
with
\begin{eqnarray}
   A(g_0) = {1 \over \Lambda^{\ind lat}} 
              \left[ b_0 g_0^2 \right]^{-{b_1 \over2 b_0^2}}
            \exp\left[ -\; {1 \over 2 b_0 g_0^2} \right] \,.
\end{eqnarray}
Note that $s(g_0)$ is the standard RG solution to the $\beta$-function.
Alternatively a second possibility would be to re-define the coupling 
constant by setting (e.g.\ \cite{Luscher:1996sc})
\begin{eqnarray}
   \tilde{g}_0^2 = g_0^2 \left\{ 1 + b_g(g_0)a\bar{m} \right\} \,,
\end{eqnarray}
and now solving
\begin{eqnarray}
   a \left. {\partial \over \partial a} \right|_{\ind physics} 
                         \tilde{g}_0 = B_0(\tilde{g}_0) \,,
\end{eqnarray}
gives
\begin{eqnarray}
   b_g(g_0) = 2 {B_0(g_0) \over g_0} \, r(g_0) 
           \underbrace{=}_{g_0^2 \to 0} -2b_{10}^{\ind lat} g_0^4 + O(g_0^6) \,.
\end{eqnarray}
As anticipated we have $b_g \propto r$. As discussed in e.g.\
\cite{Sint:1995ch,DallaBrida:2023fpl} we expect $b_g$ (and hence $r$) 
to be numerically quite small.


\section{Lattice data}

\begin{figure}[!t]
\begin{minipage}{0.45\textwidth}

      \begin{center}
        \vspace*{-0.25in}
         \includegraphics[width=6.75cm]
               {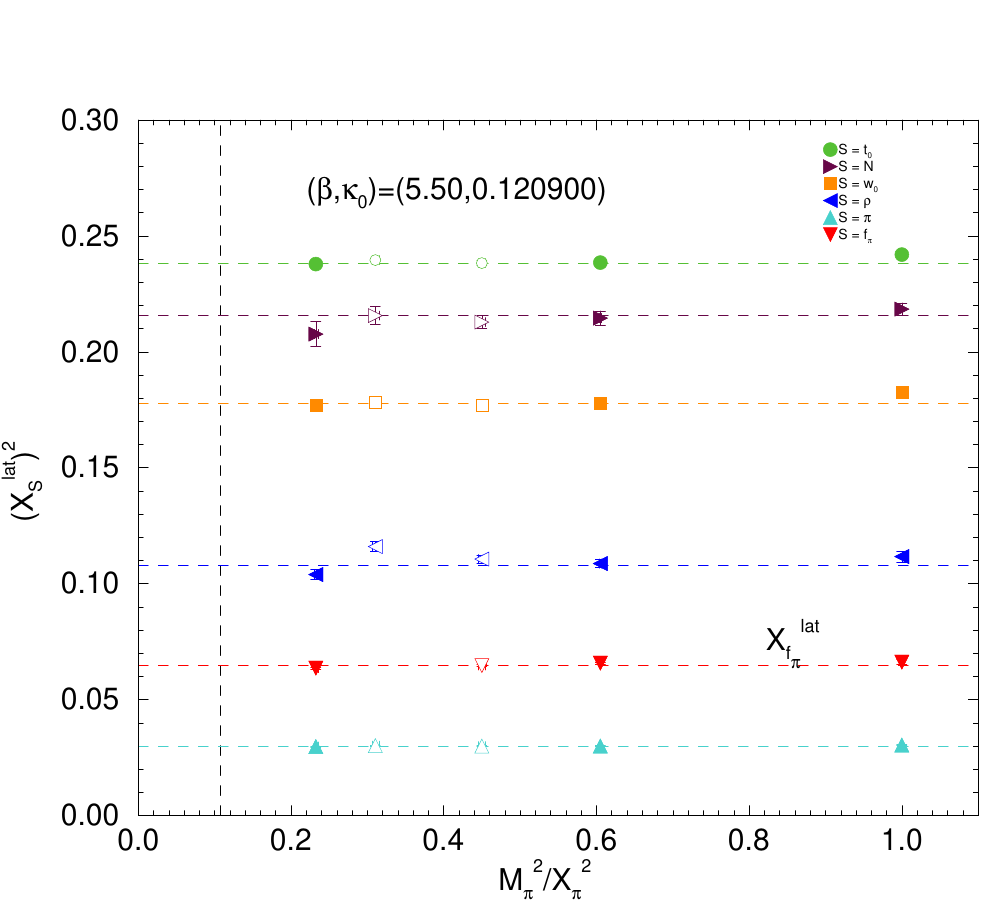}
      \end{center} 

\end{minipage}\hspace*{0.05\textwidth}
\begin{minipage}{0.50\textwidth}

      \begin{center}
         \includegraphics[width=7.00cm]
                            {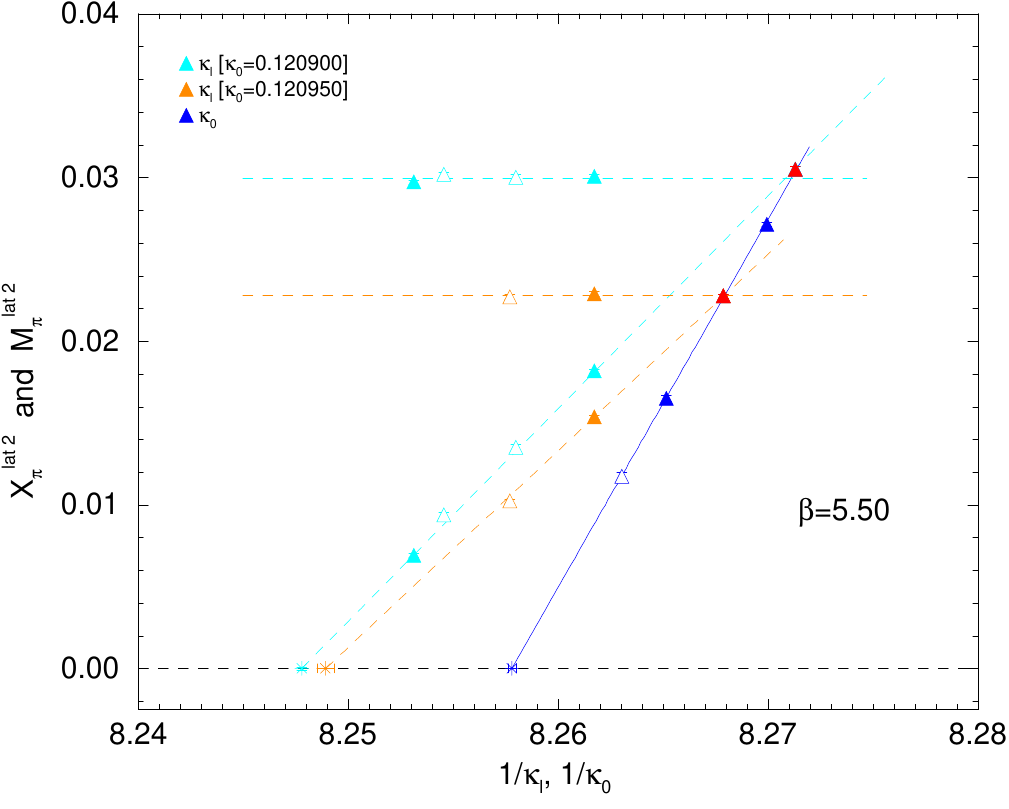}
      \end{center} 

\end{minipage}
\caption{Left panel: $X_s^{{\ind lat}\,2}$ for $s = t_0$, $N$, $w_0$, $\rho$, $\pi$,
                     $f_\pi$, for $(\beta,\kappa_0) = (5.50,0.120900)$,
                     together with constant fits. The opaque points
                     have $M_\pi L \lsim 4$ and are not considered in the
                     fit. The vertical line is approximately the physical 
                     mass ratio value. The plot is taken from 
                     \protect\cite{QCDSF-UKQCD:2016rau},
                     to which we refer to for further details.
        Right panel: $X_\pi^{{\ind lat}\,2}$ along the
                     $SU(3)$-flavour symmetric line, blue triangles
                     (with a linear fit)
                     and along $a\bar{m} = \mbox{const.}$ lines
                     starting from $\kappa_0 = 0.120900$, cyan horizontal
                     triangles and $\kappa_0 = 0.120950$, orange horizontal
                     triangles (both together with a constant fit). 
                     Also shown are the $M_\pi^{{\ind lat}\,2}$ lines
                     for constant $\kappa_0 = 0.120900$, $\kappa_0 = 0.120950$
                     and varying $\kappa_u = \kappa_s \equiv \kappa_l$,
                     cyan and orange triangles (both with a linear fit).}
\label{b5p50_mps2_ookl}
\end{figure}

For $n_f = 2+1$, $q = u$, $d$, $s$ flavours 
($am_u = am_d \equiv am_l$) the QCDSF strategy
\cite{Bietenholz:2011qq} is to consider a flavour 
singlet quantity $X_s(m^{\ind rgi}_u, m^{\ind rgi}_d, m^{\ind rgi}_s)$ for example
$X_\pi^2  = \left( 2M_K^2 + M_\pi^2 \right)/3$,
$X_{t_0}^2 = 1 / t_0$, $X_N^2  = ( M_N^2 + M_\Sigma^2 + M_\Xi^2 )/3$.
These are invariant under all $u$, $d$, $s$ quark interchanges (by definition)
and have the useful property that they have a stationary point on the
$SU(3)$-flavour symmetric line when all the quark masses are equal.
This means that
$   X_s( \bar{m}^{\ind rgi} + \delta m^{\ind rgi}_u, 
         \bar{m}^{\ind rgi} + \delta m^{\ind rgi}_d,
         \bar{m}^{\ind rgi} + \delta m^{\ind rgi}_s )
      = X_s(\bar{m}^{\ind rgi}, \bar{m}^{\ind rgi},\bar{m}^{\ind rgi})
                   + O( (\delta m^{\ind rgi}_q)^2 ) $
where $\delta m^{\ind rgi}_q$ is the `distance' from the $SU(3)$-flavour 
symmetric line, $\bar{m}^{\ind rgi}$ being held constant
(otherwise additional terms arise)
\cite{Bietenholz:2011qq}. As from eq.~(\ref{mbarrgi_mbar})
$\bar{m}^{\ind rgi} \propto a\bar{m}$, this result is valid whether 
renormalised or lattice is being considered. This is shown
in the LH panel of Fig.~\ref{b5p50_mps2_ookl}.
We use an $O(a)$ non-pertubatively improved clover action, 
\cite{Cundy:2009yy}, with $\beta = 10/g_0^2 = 5.40$, $5.50$, $5.65$, 
$5.80$ and $5.95$ on a variety of lattice sizes ($24^3\times 48$, 
$32^3\times 64$ and $48^3\times 96$) with $M_\pi L \gsim 4$.
Little evidence of any quadratic term $O((\delta(am_l))^2)$ and any 
systematic effect is seen for any $X^{{\ind lat}\,2}_s$ down to the
physical point, so to within our accuracy we shall take $X^{{\ind lat}\,2}_s$
to be constant along any $a\bar{m} = \mbox{const}$ line.
The RH panel in Fig.~\ref{b5p50_mps2_ookl} shows the ratio
$X_\pi^{{\ind lat}\,2}$ for various $\kappa_0$ and $\kappa_l$ values 
as described in the figure caption. We have
\begin{eqnarray}
   X_s^{{\ind lat}\,2}(g_0, \bar{m}^{\ind rgi})
      = a^2(g_0, \bar{m}^{\ind rgi}) X_s^2(\bar{m}^{\ind rgi}) \,,
\end{eqnarray}
(again setting $\bar{m}^{\ind rgi} \propto a\bar{m}$ in $a$) and forming a 
ratio gives
\begin{eqnarray}
   { X_\pi^{{\ind lat}\,2} \over X_{t_0}^{{\ind lat}\,2} }
      =  { X_\pi^2(\bar{m}^{\ind rgi}) 
                      \over X_{t_0}^2(\bar{m}^{\ind rgi}) }
      =  { X_\pi^{\prime 2}(0)\bar{m}^{\ind rgi} + \ldots
                      \over X_{t_0}^2(0) + \ldots }
      =  { X_\pi^{\prime 2}(0)\over X_{t_0}^2(0) } \bar{m}^{\ind rgi} + \ldots
      =  { X_\pi^{\prime 2}(0)\over X_{t_0}^2(0) } u(g_0) a\bar{m} + \ldots \,.
\label{X2oX2}
\end{eqnarray}
The first equality shows that $X_\pi^{{\ind lat}\,2}/X_{t_0}^{{\ind lat}\,2}$ 
can be taken as a proxy for $\bar{m}^{\ind rgi}$. The second (and
third) equalities use PCAC while the last equality has used the previous
relation between $\bar{m}^{\ind rgi}$ and $a\bar{m}$, eq.~(\ref{mbarrgi_mbar}).
(So, for example, determining the $1/\kappa_0$ gradient of the $SU(3)$-flavour 
symmetric line could give an indication of 
$(X_\pi^{\prime 2}(0) / X_{t_0}^2(0)) u(g_0)$.)
However, while the RG equations in section~\ref{RG_eqn} were developed with
$a\bar{m}$, the present long extrapolation in $1/\kappa_0$ necessary to 
find $\kappa_{0c}$ means that it is difficult to determine $a\bar{m}$
and so from eq.~(\ref{X2oX2}) we shall use instead
$X_\pi^{{\ind lat}\,2} / X_{t_0}^{{\ind lat}\,2}$.


\section{Lattice results}

\begin{figure}[!t]
\begin{minipage}{0.45\textwidth}

      \begin{center}
         \includegraphics[width=6.75cm]
             {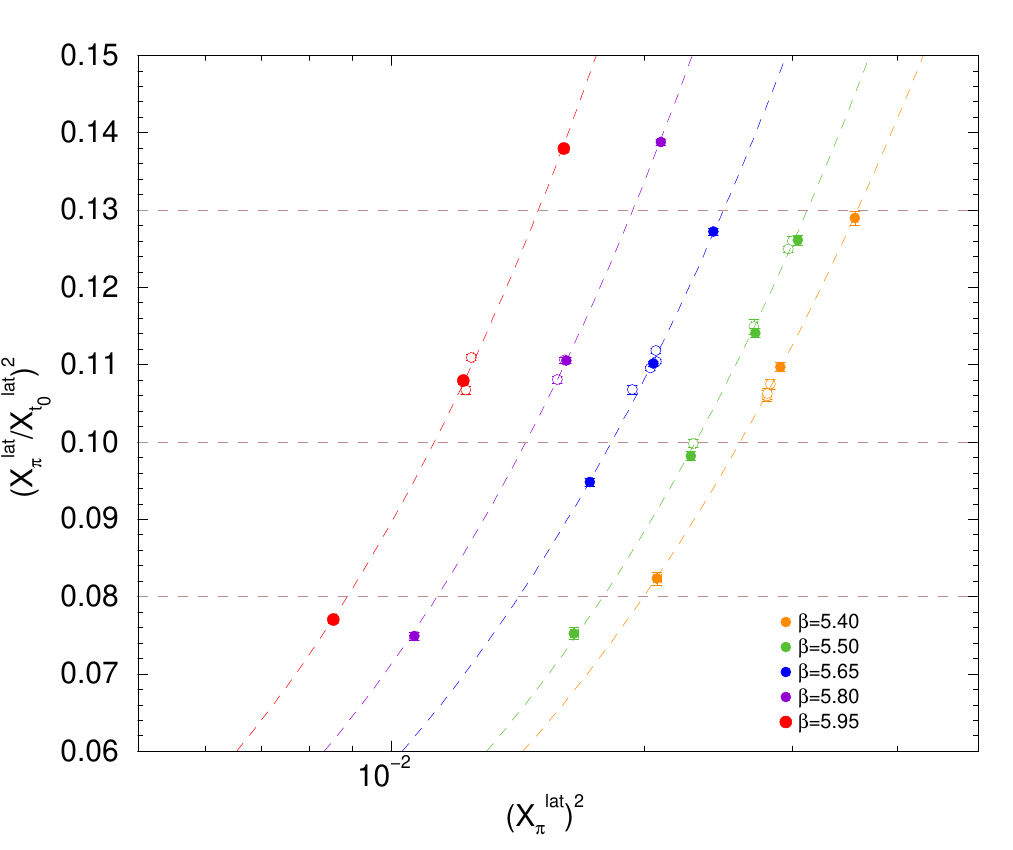}
      \end{center} 

\end{minipage}\hspace*{0.05\textwidth}
\begin{minipage}{0.50\textwidth}

      \begin{center}
         \includegraphics[width=7.25cm]
                        {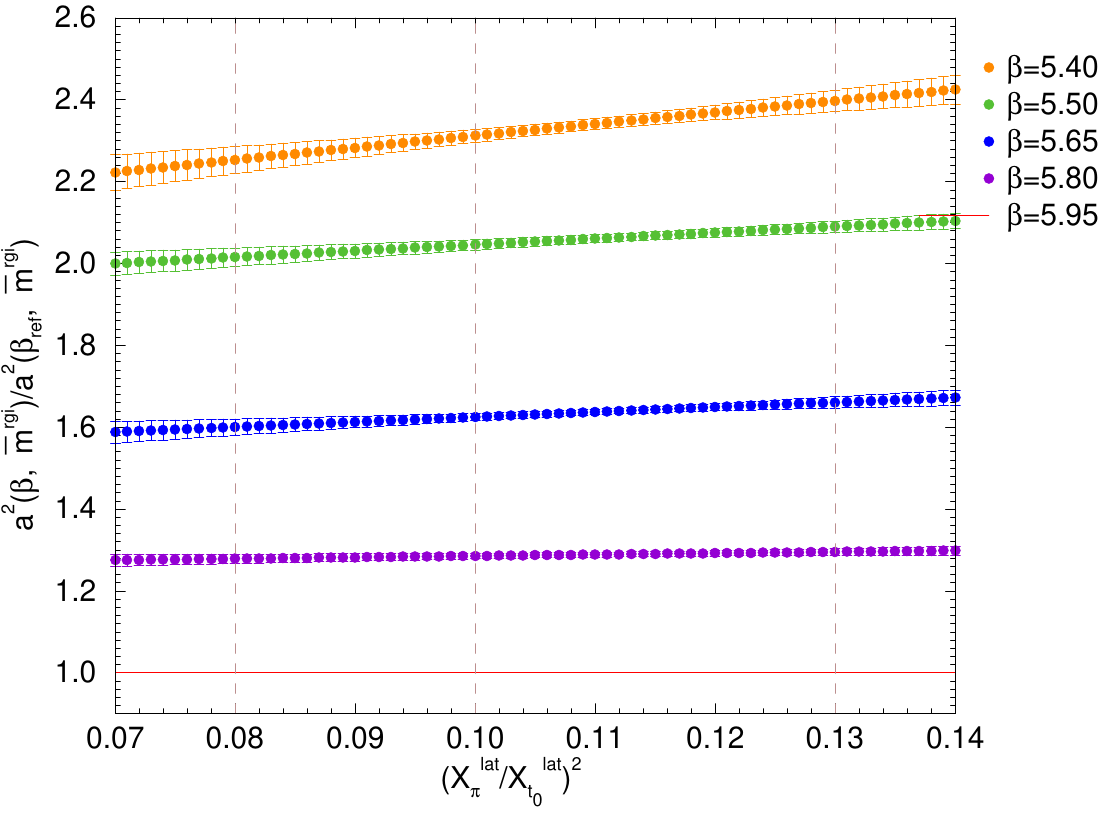}
      \end{center} 

\end{minipage}
\caption{Left panel: $y = X^{{\ind lat}\,2}_\pi/X^{{\ind lat}\,2}_{t_0}$ versus 
                     $x = X^{{\ind lat}\,2}_\pi$ (logarithmic scale) together 
                     with a $2$-parameter fit 
                     $x = y(A + By)$ 
                     for our five $\beta$ values. 
                     (the filled points lie on the $SU(3)$-flavour 
                     symmetric lines, the opaque points are 
                     $a\bar{m}=\mbox{const.}$).
                     Sample lines when the 
                     $y$-axis height $y_0 = 0.08$, $0.10$, $0.13$.
           Right panel: $a^2(\bar{m}^{\ind rgi},\beta)
                        /a^2(\bar{m}^{\ind rgi},\beta_{\ind ref})$ 
                     with $\beta_{\ind ref} = 5.95$ versus 
                     $X^{{\ind lat}\,2}_\pi/X^{{\ind lat}\,2}_{t_0}$ for our five 
                     $\beta$ values. The vertical lines correspond to
                     the sample horizontal lines in the LH panel.}
\label{b5pxy_Xpi2oXosqrtt02_a2Xpi2_rev_axes_loglin_3parm}
\end{figure}

We now compare lattice spacings as a function of $\beta$ with
matching physics (i.e.\ same $\bar{m}^{\ind rgi}$). So we plot for the $y$- 
and $x$-axes
\begin{eqnarray}
   y = {X_\pi^{{\ind lat}\,2} \over X_{t_0}^{{\ind lat}\,2}}
     = {X_\pi^2(\bar{m}^{\ind rgi}) \over 
          X_{t_0}^2(\bar{m}^{\ind rgi})} \,, \qquad
   x = X_\pi^{{\ind lat}\,2}
          = a^2(g_0, \bar{m}^{\ind rgi}) X_\pi^2(\bar{m}^{\ind rgi}) \,.
\end{eqnarray}
The $y$-axis is a proxy for the (physical) quark mass, $\bar{m}^{\ind rgi}$.
In the LH panel of 
Fig.~\ref{b5pxy_Xpi2oXosqrtt02_a2Xpi2_rev_axes_loglin_3parm}
we show this plot, together with a $2$-parameter fit
$x = y(A + By)$, and with sample lines with the 
$y$-axis height $y_0 = 0.08$, $0.10$, $0.13$.
The physical region is given by $y_0 \sim 0.09$ -- $0.10$.

If $a$ depends only on $\beta$ and negligibly on $\bar{m}^{\ind rgi}$
the lines will have the same slope at each point and be equidistant 
to each other. 
This can be illustrated by considering two $\beta$-values ($\beta$ and 
$\beta_{\ind ref}$ say) with common $\bar{m}^{\ind rgi}$ (i.e.\ the same height 
for $y$-axis) then the ratio of the corresponding values on the $x$-axis 
is the ratio of lattice spacings only
\begin{eqnarray}
   {x(\beta, \bar{m}^{\ind rgi}) \over
        x(\beta_{\ind ref}, \bar{m}^{\ind rgi})}
     = { a^2(\beta, \bar{m}^{\ind rgi})
                \over a^2(\beta_{\ind ref}, \bar{m}^{\ind rgi}) } \,,
\label{a2oa2}
\end{eqnarray}
which when re-written as a difference:
$\ln x(\beta, \bar{m}^{\ind rgi}) 
      - \ln x(\beta_{\ind ref}, \bar{m}^{\ind rgi})
         = \ln \left( a^2(\beta, \bar{m}^{\ind rgi})
                      / a^2(\beta_{\ind ref}, \bar{m}^{\ind rgi}) \right)$
should be approximately constant. This appears to be the case.

To check this in the RH panel of
Fig.~\ref{b5pxy_Xpi2oXosqrtt02_a2Xpi2_rev_axes_loglin_3parm} we plot 
$a^2(\beta,\bar{m}^{\ind rgi})/a^2(\beta_{\ind ref},\bar{m}^{\ind rgi})$
with $\beta_{\ind ref} = 5.95$ against $X_\pi^{{\ind lat}\,2}/X_{t_0}^{{\ind lat}\,2}$
and thus for many $\bar{m}^{\ind rgi}$ values%
\footnote{The errors are estimated by considering fits
          $x = y(A + B(y-y_0))$ and varying $y_0$.
          The error in $A$ gives us the error in $x$ at
          $y = y_0$, then used in eq.~(\ref{a2oa2}).}.
If there was no $\bar{m}^{\ind rgi}$ dependence these
would be constant. Only for the coarsest lattice ($\beta = 5.40$) is there 
an appreciable deviation from constant behaviour, and this is small.
For example in the figure range a $10\%$ change in 
$X_\pi^2(\bar{m}^{\ind rgi})/X_{t_0}^2(\bar{m}^{\ind rgi})$ gives 
$\sim 1$-$2\%$ change in 
$a^2(5.40,\bar{m}^{\ind rgi})/a^2(\beta_{\ind ref},\bar{m}^{\ind rgi})$.
So effectively there is only an overall shift in scale
if we slightly miss a chosen value of the quark mass, 
$\bar{m}^{{\ind rgi}\,*}$.


\section{Lattice scaling}


As the lattice spacing data is well described by linear behaviour in the
quark mass, we can apply the formalism developed in the previous sections
to investigate the scaling behaviour. From eq.~(\ref{a_definition}) we can
form $a^2(\beta,\bar{m}^{\ind rgi})/a^2(\beta_{\ind ref},\bar{m}^{\ind rgi})$.
This gives
\begin{eqnarray}
   {a^2(\beta,\bar{m}^{\ind rgi}) \over 
            a^2(\beta_{\ind ref},\bar{m}^{\ind rgi}) }
   = {s^2(\beta) \over s^2(\beta_{\ind ref})}
      \left\{ 1 + D_{\pi/t_0}(\beta,\beta_{\ind ref})
                      {X_\pi^2(\bar{m}^{\ind rgi}) \over
                                     X_{t_0}^2(\bar{m}^{\ind rgi})}
      \right\} \,.
\label{a2oa2_full}
\end{eqnarray}
with
\begin{eqnarray}
   D_{\pi/t_0}(\beta,\beta_{\ind ref})
      = 2 \left( 
         {r(\beta) \over u(\beta)} -{r(\beta_{\ind ref}) \over u(\beta_{\ind ref})}
        \right) {X_{t_0}^2(0) \over X_\pi^{\prime 2}(0)} \,.
\end{eqnarray}
From our previous fits (LH panel of 
Fig.~\ref{b5pxy_Xpi2oXosqrtt02_a2Xpi2_rev_axes_loglin_3parm} and the
associated fit parameters $A$, $B$) we can determine
$s^2(\beta) / s^2(\beta_{\ind ref})$ and the coefficient of
$X_\pi^2(\bar{m}^{\ind rgi}) / X_{t_0}^2(\bar{m}^{\ind rgi})$.

First recalling the solution for $s(g_0)$, eq.~(\ref{sg0}), we have
\begin{eqnarray}  
   { s^2(g_0) \over s^2(g_{0\,{\ind ref}}) }
      = \exp\left[ -2\int^{g_0}_{g_{0\,{\ind ref}}} {d\xi \over B_0(\xi)} \right] \,.
\label{ratio_beta}
\end{eqnarray}
Guided by the knowledge of the $\beta$-function as given in 
eq.~(\ref{PT_funs}) we use the simplest $[2/2]$ Pad{\'e} approximation
as a fit function
\vspace*{-0.125in}
\begin{eqnarray}
   B_0(g_0) = - \beta_{[2/2]}(g_0)
           = b_0 g_0^3 \, { 1 +
             \left( {b_1\over b_0} - {b_2^{\ind eff}\over b_1} \right) g_0^2 \over
                      1 - {b_2^{\ind eff}\over b_1} g_0^2 } \,.
\label{B0_pade}
\end{eqnarray}
(Note that eq.~(\ref{ratio_beta}) can be integrated analytically.) 
Secondly we have
\begin{eqnarray}
   D_{\pi/t_0}(\beta, \beta_{\ind ref})
   = 2 {X_{t_0}^2(0) \over X_\pi^{\prime 2}(0)}
     \int_{g_{0\,{\ind ref}}}^{g_0} d\xi { B_1(\xi) \over B_0^2(\xi)u(\xi) }
    \,\, \underbrace{\propto}_{g_0^2\to 0} \,\, 
           \int_{1/(2b_0g^2_{0\,\ind ref})}^{1/(2b_0g_0^2)} d\xi \,\xi^a e^{-\xi} \,,
\label{roDmroD_fit}
\end{eqnarray}
where the LO result ($g_0^2 \to 0$) is given by the last expression 
with $a=(b_1+b_0d_0)/(2b_0^2)$ and can be written as the difference of 
incomplete-$\Gamma$ functions. Numerically in our region of interest
it is almost linear in $\beta$: i.e.\ $\propto \beta_{\ind ref}-\beta$.

\begin{figure}[!t]
\begin{minipage}{0.45\textwidth}

      \begin{center}
         \includegraphics[width=6.75cm]
           {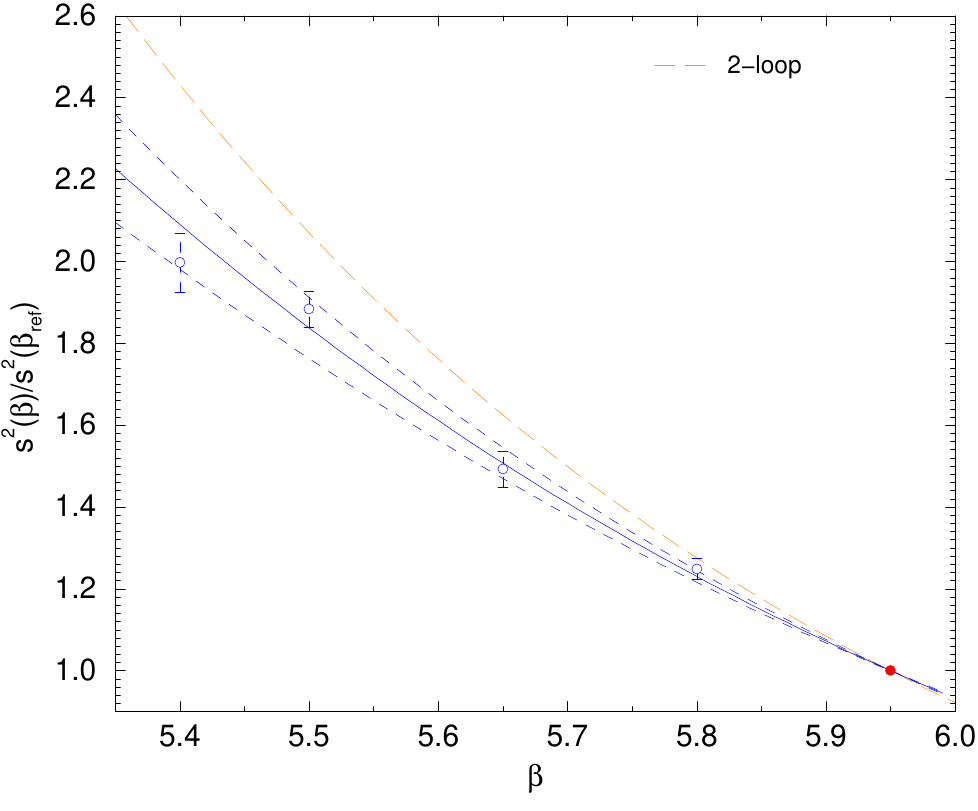}
      \end{center} 

\end{minipage}\hspace*{0.05\textwidth}
\begin{minipage}{0.50\textwidth}

      \begin{center}
         \includegraphics[width=6.75cm]
                           {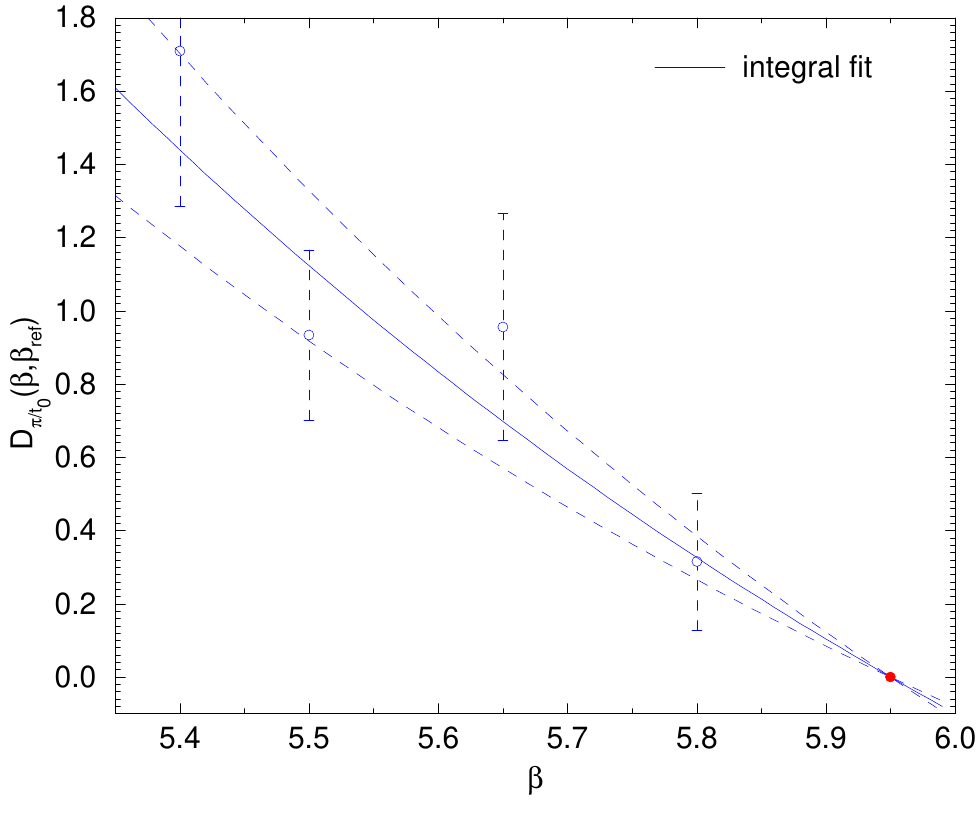}
      \end{center} 

\end{minipage}
\caption{Preliminary results. Left panel: $s^2(\beta)/s^2(\beta_{\ind ref})$
            versus $\beta$ ($\beta_{\ind ref} = 5.95$) blue line from 
            the fit results given by the ratios from
            eq.~(\protect\ref{ratio_beta}) (with eq.~(\protect\ref{B0_pade}))
            as the blue points. 
            The orange dashed line is the result from the two-loop 
            $\beta$-function ($b_2^{\ind eff} = 0$).
         Right panel: $D_{\pi/t_0}(\beta,\beta_{\ind ref})$ versus $\beta$, 
            with the fit as indicated in eq.~(\ref{roDmroD_fit}).}
\label{beta_ratio_x0_Xpi2oXosqrtt02_a2Xpi2}
\end{figure}

In the LH panel of  Fig.~\ref{beta_ratio_x0_Xpi2oXosqrtt02_a2Xpi2}
we show $s^2(\beta)/s^2(\beta_{\ind ref})$ as blue points (from the
previous fit), together with a further fit using eqs.~(\ref{ratio_beta}), 
(\ref{B0_pade}). Also shown is the two-loop $\beta$-function result 
(i.e.\ setting $b_2^{\ind eff} = 0$ in eq.~(\ref{B0_pade})) as an orange 
dashed line. $s^2(\beta)/s^2(\beta_{\ind ref})$ is small, although growing
at $\beta = 5.40$, it is still only about $\sim 10\%$.
The RH panel of Fig.~\ref{beta_ratio_x0_Xpi2oXosqrtt02_a2Xpi2} shows
$D_{\pi/t_0}(\beta,\beta_{\ind ref})$ together with a fit as suggested by 
eq.~(\ref{roDmroD_fit}) (with a one parameter fit for the proportionality
constant). 

\begin{figure}[!t]
   \begin{center}
      \includegraphics[width=7.50cm]{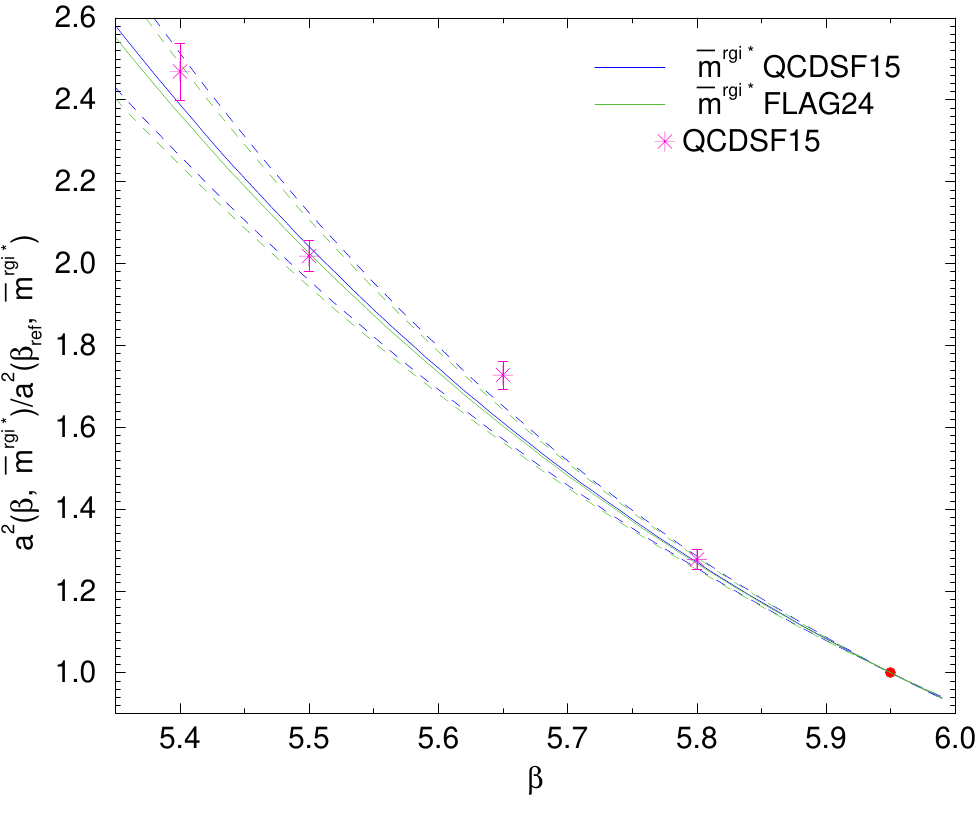}
   \end{center} 
\caption{Preliminary results. 
         $a^2(\beta,\bar{m}^{{\ind rgi}\,*})
                      / a^2(\beta_{\ind ref},\bar{m}^{{\ind rgi}\,*})$
         against $\beta$ ($\beta_{\ind ref} = 5.95$),
         using our previous determination of 
         $X_\pi^2(\bar{m}^{{\ind rgi}\,*})/X_{t_0}^2(\bar{m}^{{\ind rgi}\,*})$ ,
         QCDSF15 \protect\cite{Bornyakov:2015eaa}, as a blue line.
         Also shown are the results arising from FLAG24 
         \protect\cite{FlavourLatticeAveragingGroupFLAG:2024oxs}
         for $X_\pi^2(\bar{m}^{{\ind rgi}\,*})/X_{t_0}^2(\bar{m}^{{\ind rgi}\,*})$,
         green line.
         Furthermore as a comparison we show our previous results, 
         QCDSF15, as magenta stars.}
\label{beta_ratio_lat24+qcdsf}
\end{figure}

Finally we wish to re-construct 
$a^2(\beta,\bar{m}^{{\ind rgi}\,*})/a^2(\beta,\bar{m}^{{\ind rgi}\,*})$,
via eq.~(\ref{a2oa2_full}), with the fit function 
results as shown in Fig.~\ref{beta_ratio_x0_Xpi2oXosqrtt02_a2Xpi2}.
In Fig.~\ref{beta_ratio_lat24+qcdsf}
we plot $a^2(\beta,\bar{m}^{{\ind rgi}\,*})
                / a^2(\beta_{\ind ref},\bar{m}^{{\ind rgi}\,*})$ 
against $\beta$ from eq.~(\ref{a2oa2_full}), using values of 
$X^2_\pi(\bar{m}^{{\ind rgi}\,*})/X^2_{t_0}(\bar{m}^{{\ind rgi}\,*}) = 0.099(3)$, 
QCDSF15 \cite{Bornyakov:2015eaa}%
\footnote{The $\beta = 5.95$ result was determined later, using the
methods described in \cite{Bornyakov:2015eaa} as $0.0521(4)\,\mbox{fm}$.}
and as a comparison $0.091(1)$, FLAG24 
\cite{FlavourLatticeAveragingGroupFLAG:2024oxs}.
As expected from our previous results (as exemplified by 
Fig.~\ref{b5pxy_Xpi2oXosqrtt02_a2Xpi2_rev_axes_loglin_3parm}) there is
little numerical difference between the two values of $\bar{m}^{{\ind rgi}\,*}$.
Comparing the results using this RG-guided approach with each value
being found separately (the magenta points), we have achieved a smoother
behaviour for the lattice spacing.


\section{Conclusions}


In this talk we have discussed RG equations for $2+1$ clover fermions
and attempted to construct some RG functions.
Considering the lattice spacing, we find scaling, with little dependence 
on $\bar{m}^{{\ind rgi}\,*}$. The next steps, \cite{qcdsf25}, could be to 
reduce the errors on the lattice spacing ratios with more $SU(3)$-flavour 
symmetric data, using a more sophisticated fit procedure and to determine 
$\bar{m}^{{\ind rgi}\,*}$ accurately at one value of $\beta_{\ind ref}$ 
and hence a value of $a\,\mbox{fm}$, using (e.g.\ ) the nucleon baryon octet, 
\cite{Bornyakov:2015eaa}. A possible advantage of the method is that lattice 
spacing ratios can be more precisely determined at many $\beta$ values 
(using e.g.\, as here, accurate pion mass and gradient flow data). 
The general possibility is that this might lead to smoother continuum 
extrapolations.


\acknowledgments


The numerical configuration generation (using the BQCD lattice QCD 
program \cite{Haar:2017ubh}) and data analysis (using the CHROMA software 
library \cite{Edwards:2004sx}) was carried out on the
DiRAC Blue Gene Q and Extreme Scaling Service
(Edinburgh Centre for Parallel Computing (EPCC), Edinburgh, UK), 
the Data Intensive Service 
(Cambridge Service for Data-Driven Discovery, CSD3, Cambridge, UK),
the Gauss Centre for Supercomputing (GCS) supercomputers JUQUEEN and JUWELS 
(John von Neumann Institute for Computing, NIC, Jülich, Germany) 
and resources provided by the North-German Supercomputer Alliance
(HLRN), the National Computer Infrastructure 
(NCI National Facility in Canberra, Australia supported by the
Australian Commonwealth Government) 
and the Phoenix HPC service (University of Adelaide). 
K.U.C. is supported by the ARC Grant No. DP220103098. 
R.H. is supported in part by the STFC Grant No. ST/X000494/1.
P.E.L.R. is supported in part by the STFC Grant No. ST/G00062X/1.
G.S. is supported by DFG Grant No. SCHI 179/8-1.
R.D.Y. and J.M.Z. are supported by the ARC Grants
No. DP220103098 and DP240102839. For the purpose
of open access, the authors have applied a Creative
Commons Attribution (CC BY) licence to any author
accepted manuscript version arising from this submission.



\end{document}